\documentstyle[preprint,prl,aps,epsf]{revtex}
\begin{document}
\draft
\tighten

\title{\hfill {\small DOE-ER-40757-103} \\
\hfill {\small UTEXAS-HEP-97-17} \\
\hfill  {\small MSUHEP-70815} \\
\hfill {\small MADPH-97-1010} \\
\hfill \\ 
Effective Lagrangians and low energy \\ 
photon-photon scattering} 

\author{Duane A. Dicus$^1$, Chung Kao$^2$ and Wayne W. Repko$^3$} 
\address{$^1$Center for Particle Physics and Department of Physics,\\
University of Texas, Austin, Texas 78712 \\
$^2$Department of Physics, University of Wisconsin, Madison, Wisconsin 53706 \\
$^3$Department of Physics and Astronomy,\\
Michigan State University, East Lansing, Michigan 48824}
\date{\today}
\maketitle

\begin{abstract}
We use the behavior of the photon-photon scattering for photon energies $\omega$
less than the electron mass, $m_e$, to examine the implications of treating the 
Euler-Heisenberg Lagrangian as an effective field theory. Specifically, 
we determine the $\omega^2/m_e^2$ behavior of the scattering amplitude 
predicted by including one-loop corrections to the Euler-Heisenberg effective 
Lagrangian together with the counterterms required by renomalizability.
This behavior is compared with the energy dependence obtained by expanding the
exact QED photon-photon scattering amplitude. If the introduction of
counterterms in the effective field theory is restricted to those determined by
renormalizability, the $\omega^2/m_e^2$ dependences of the two expansions
differ.
\end{abstract}
\pacs{11.10.Ge,14.70.Bh,21.30.Fe}

\section {Introduction}

Low energy photon-photon scattering is a textbook example of a process which can
be described using an effective interaction, the Euler-Heisenberg effective
Lagrangian \cite{Eul}. The precise form of this Lagrangian (given below) can be
obtained by expanding the box diagrams for photon-photon scattering in powers 
of the photon energy over the electron mass, $\omega/m_e$, and keeping the first
non-vanishing term, which is of order $\omega^4/m_e^4$. This leads to the 
cross section
\begin{equation}\label{sig0}
\frac{d\sigma}{d\Omega} = \frac {139\alpha^4}{(180\pi)^2}\,
(3+z^2)^2\frac{\omega^6}{m_e^8}\;,
\end{equation}
with its characteristic $\omega^6$ dependence on the photon energy. Here, $z$ 
is the cosine of the scattering angle. The energy scale in this case is $m_e$,
a result of `integrating out' the heavy degree of of freedom, the
electron, to obtain an effective interaction. 

There are compelling arguments suggesting that effective interactions can be 
used to define effective field theories which are adequate descriptions of 
Nature if one accepts a reasonable set of assumptions about the 
renormalization program \cite{wein}. With these assumptions, it is possible to
obtain finite corrections to physical processes from the (ordinarily
divergent) contributions calculated by using the effective interactions in 
higher orders of perturbation theory. The price for extracting these finite
corrections is the introduction of local counterterms whose couplings are not
known {\em a priori}. It is, however, clear that the counterterms introduce
corrections of higher order in the ratio of the energy to the energy scale.

The implications of this approach can be explored using the laboratory of
photon-photon scattering as is evident from the plot of the total cross section
given in Fig.\,1 \cite{Kar,DeT}. Here, it can be seen that the $\omega^6$
behavior persists up to $\omega\simeq$ 0.4 $m_e$ or so, at which point the slope
increases indicating the presence of higher powers of $\omega$. Our purpose 
is to compute the one-loop corrections to low energy photon-photon scattering
amplitude using the Euler-Heisenberg effective Lagrangian with the appropriate 
counterterms and to compare this result with that obtained by expanding the
{\em exact} amplitude in powers of $\omega/m_e$. This will enable us to better
understand how an effective field theory mimics a full field theory in
a situation where explicit calculations in both approaches are feasible.

\section{One-loop corrections}\label{loop}

The full Euler-Heisenberg Lagrangian describing photon self interactions is an
expansion in powers of the electromagnetic field tensor $F_{\mu\nu}$. Assuming
parity conservation, it can be expressed in terms of the two independent, gauge
invariant combinations $F_{\mu\nu}F^{\mu\nu}$ and
$F_{\mu\nu}F^{\nu\lambda}F_{\lambda\rho}F^{\rho\mu}$. Due to the antisymmetry of
$F_{\mu\nu}$, scalars formed from a odd number of field tensors vanish.
For a one-loop calculation, it suffices to use the terms 
\begin{eqnarray} \label{E-H}
{\cal L}_{\rm E-H} & = & \frac{1}{180}\frac{\alpha^2}{m_e^4}
\left[5\,\left(F_{\mu\nu}F^{\mu\nu}\right)^2 - 
14\,F_{\mu\nu}F^{\nu\lambda}F_{\lambda\rho}F^{\rho\mu}\right] \nonumber \\
&   &+ \frac{1}{315}\frac{\pi\alpha^3}{m_e^8}\left[9\;
\left(F_{\mu\nu}F^{\mu\nu}\right)^3 
- 26\;F_{\mu\nu}F^{\nu\lambda}F_{\lambda\rho}F^{\rho\mu}
F_{\alpha\beta}F^{\alpha\beta}\right]\,.
\end{eqnarray}
The one-loop contribution of the order $\alpha^3$ term in Eq.\,(\ref{E-H})
to the photon-photon scattering amplitude, obtained by
contracting a pair of $F_{\mu\nu}$'s, actually vanishes by dimensional 
regularization \cite{2to4}. This leaves only the $\alpha^4$ one-loop 
contribution from the
first term of Eq.\,(\ref{E-H}) used in the second order of perturbation theory.
Because the photons are massless, all the loop integrals, which take the form
\begin{equation}
B_{\alpha_1\cdots\alpha_i} = \frac{1}{i\pi^2}\int\,d^{\,4}q\,\frac{
q_{\alpha_1}\cdots q_{\alpha_i}}{q^2(q + k)^2}\,,
\end{equation}
$i = 1,\cdots, 4$, are proportional to the scalar integral 
\begin{equation}
B_0(k^2) = \frac{1}{i\pi^2}\int\,d^{\,4}q\,\frac{1}{q^2(q + k)^2}\,,
\end{equation}
multiplied by a polynomial in the momentum $k$. Evaluating $B_0(k^2)$ using a 
cutoff $\Lambda$, we find
\begin{equation}
B_0(k^2) = \ln\left(\frac{\Lambda^2}{-k^2 - i\varepsilon}\right)\,.
\end{equation}

In second order, the $\alpha^2$ term in Eq.\,(\ref{E-H}) has $s$, $t$ and $u$
channel contributions, illustrated in Fig.\,2. As a consequence, the
one-loop correction to the helicity amplitudes can be written
\begin{equation}\label{amps}
T^{\,(2)}_{\lambda_1\lambda_2\lambda_3\lambda_4}  = \frac{\alpha^4}{(45\pi)^2}
\frac{\omega^8}{m_e^8}\left(T^s_{\lambda_1\lambda_2\lambda_3\lambda_4}B_0(s) +
T^t_{\lambda_1\lambda_2\lambda_3\lambda_4}B_0(t) +
T^u_{\lambda_1\lambda_2\lambda_3\lambda_4}B_0(u)\right)\,,
\end{equation}
with $s = (k_1 + k_2)^2$, $t = (k_1 - k_3)^2$ and $u = (k_1 - k_4)^2$. Since the
vertex functions resulting from ${\cal L}_{\rm E-H}$ are quite complicated
\cite{hal}, the programs SCHOONSCHIP \cite{velt} and FORM \cite{verm} were used
to perform the algebraic manipulations. The
explicit forms of the $T^i_{\lambda_1\lambda_2\lambda_3\lambda_4}$ are given in
Table I. In order to isolate the cutoff dependence, we introduce a
renormalization scale $\mu$ and rewrite Eq.\,(\ref{amps}) as
\begin{equation}\label{renamp}
T^{\,(2)}_{\lambda_1\lambda_2\lambda_3\lambda_4}  = \frac{\alpha^4}{(45\pi)^2}
\frac{\omega^8}{m_e^8}\ln\left(\frac{\Lambda^2}{\mu^2}\right)\,\sum_{i = s,t,u}
T^i_{\lambda_1\lambda_2\lambda_3\lambda_4} + 
\;\tilde{T}^{\,(2)}_{\lambda_1\lambda_2\lambda_3\lambda_4}\,,
\end{equation}
where $\tilde{T}^{\,(2)}_{\lambda_1\lambda_2\lambda_3\lambda_4}$ is obtained 
from Eq.\,(\ref{amps}) by replacing $\Lambda^2$ with $\mu^2$ in $B_0$. The sums 
of the various helicity combinations are given in the last row of Table I.

To complete the interpretation of the cutoff-dependent terms in Eq.
(\ref{renamp}) as contributions to coupling constant renormalization, it is
necessary to show that they can be obtained as matrix elements of local, gauge
invariant operators. In addition to four $F_{\mu\nu}$'s, these operators must
contain four derivatives in order to produce the $\omega^8$ behavior.
Finding them is something of a trial and error process. There 
are many tensors which can be formed with the required properties, but most of 
them are not independent. The forms given below are sufficient to reproduce 
the helicity amplitudes in the last row of Table I.
\begin{eqnarray} \label{counter}
{\cal L}_{\rm E-H}^{(2)} &=&\frac{1}{10(90\pi)^2}\frac{\alpha^4}{m_e^8}
\ln\left(\frac{\Lambda^2}{\mu^2}\right)
\left[182\left(\partial_{\alpha}F_{\mu\nu}\right)\!\left(\partial^
{\alpha}F^{\mu\nu}\right)\!\left(\partial_{\beta}F_{\lambda\rho}\right)\!
\left(\partial^{\beta}F^{\lambda\rho}\right)\right. \nonumber \\
& &\left.-639\left(\partial_{\alpha}
F_{\mu\nu}\right)\!\left(\partial^{\beta}F^{\mu\nu}\right)\!\left(\partial^
{\alpha}F_{\lambda\rho}\right)\!\left(\partial_{\beta}F^{\lambda\rho}\right)
+ 1523\left(\partial^{\alpha}\partial_{\beta}
F_{\mu\lambda}\right)F^{\lambda\nu}\left(\partial^{\mu}\partial_{\nu}
F_{\alpha\rho}\right)F^{\rho\beta}\right].
\end{eqnarray}

The effective Lagrangian
\begin{eqnarray} \label{lcount}
{\cal L} & = & {\cal L}_{\rm E-H} + \frac{1}{m_e^8}\left[a_1
\left(\partial_{\alpha}F_{\mu\nu}\right)\left(\partial^{\alpha}F^{\mu\nu}\right)
\left(\partial_{\beta}F_{\lambda\rho}\right)\left(\partial^{\beta}
F^{\lambda\rho}\right)\right. \nonumber \\
&   &\left.+ a_2\left(\partial_{\alpha}F_{\mu\nu}\right)\left(\partial^{\beta}
F^{\mu\nu}\right)\left(\partial^{\alpha}F_{\lambda\rho}\right)
\left(\partial_{\beta}F^{\lambda\rho}\right)
+ a_3\left(\partial^{\alpha}\partial_{\beta}F_{\mu\lambda}\right)F^{\lambda\nu}
\left(\partial^{\mu}\partial_{\nu}F_{\alpha\rho}\right)F^{\rho\beta}\right]
\end{eqnarray}
then gives a finite matrix element for elastic scattering at the one-loop level
provided the cutoff dependent terms represented by 
Eq.\,(\ref{counter}) are interpreted as renormalizations of
the coupling constants $a_1, a_2$\, and $a_3$.

\section{Corrections to the QED Amplitude}\label{qed}

The leading $\omega^6$ dependence of the low energy cross section follows from 
the fact that the terms in Eq.\,(\ref{E-H}) involving four field tensors are of
dimension 8. Thus, the amplitudes go as $\omega^4$ and, from QED or using
Eq.\,(\ref{E-H}), one can obtain the helicity amplitudes \cite{Kar} given in 
the first row of 
Table II. The remaining amplitudes are equal to these by parity and time 
reversal. Equation\,(\ref{sig0}) follows by taking the sum of the squares of 
the $T^4_{\lambda_1\lambda_2\lambda_3\lambda_4}$.

Table III compares the total cross section, as calculated approximately by
integrating Eq.\,(\ref{sig0}) over $z$,
\begin{equation}\label{xsec4}
\sigma =\frac{1}{(45)^2}\;\frac{973}{5}\,\frac{\alpha^4}{\pi}\;
\frac{\omega^6}{m_e^8}\,,
\end{equation}
with the full one loop QED expression calculated by using a rapidly converging
series for the Spence functions \cite{Kao} and performing a numerical 
integration over $z$.  As can be seen, Eq.\,(\ref{xsec4}) provides a 
reasonable approximation for $\omega/m_e \alt 0.50$.

Again, corrections to Eqs.\,(\ref{sig0}) or (\ref{xsec4}) can only involve 
effective interactions with
derivatives acting on four factors of $F_{\mu\nu}$ and, by Lorentz invariance,
must involve an even number of derivatives.  Thus the helicity amplitudes and
therefore the cross section are expansions in $\omega^2/m_e^2$.  
For comparison purposes, we have
calculated the terms of order $\omega^6$ and $\omega^8$ by expanding the
electron box diagrams of QED. The results are given in the second and third rows
of Table II. 

These amplitudes add terms to Eq.\,(\ref{sig0}) which becomes
\begin{eqnarray}
\frac{d\sigma}{d\Omega}&=&\frac{\alpha^4}{(180\pi)^2}\,
\frac{\omega^6}{m_e^8} \left[ 139\,(3+z^2)^2+ 
160\frac{\omega^2}{m_e^2}\, (1-z^2)(3+z^2)\right. \nonumber \\
& &\left. +\frac{4}{245}\frac{\omega^4}{m_e^4}\;\left(48533 + 35885\,z^2 
+ 18995\,z^4 + 1641\,z^6\right)\right]\,,
\end{eqnarray}
where the first term is the contribution from the square of the $\omega^4$
amplitude, the second term is the cross term between the amplitudes of
order $\omega^4$ with those of order $\omega^6$, the third term consists of
the square of the $\omega^6$ amplitudes and the cross terms between the 
$\omega^4$ amplitudes and those of order $\omega^8$. The
corresponding total cross section is
\begin{equation}\label{xsec6}
\sigma = \frac{1}{(45)^2}\;\frac{\alpha^4\omega^6}{\pi m_e^8}\;\left[
\;\frac{973}{5} + \frac{128}{3}\; \frac{\omega^2}{m_e^2}
+ \frac{409792}{1715}\;\frac{\omega^4}{m_e^4}\right]\,.
\end{equation}
The last column of Table III gives the ratio of Eq.\,(\ref{xsec6}) to the full 
one loop cross section.  Inclusion of the second and third terms in 
Eq.\,(\ref{xsec6}) provides a very good approximation, all the way to 
$\omega/m_e \simeq 0.90$.

As mentioned above, the helicity amplitudes in the first row of Table II can be
obtained from the effective Lagrangian
\begin{equation}\label{eh}
{\cal L}_{\rm eff} = \frac{1}{180}\frac{\alpha^2}{m_e^4}\left[
5\,\left(F_{\mu\nu}F^{\mu\nu}\right)^2 - 
14\,F_{\mu\nu}F^{\nu\lambda}F_{\lambda\rho}F^{\rho\mu}\right]\,.
\end{equation}
In order to generalize this expression to reproduce the $\omega^6$ and
$\omega^8$ corrections, it is necessary to construct terms containing two or
four additional derivatives acting on a combination of four
$F_{\mu\nu}$'s to produce a scalar. Terms with four derivatives were encountered
in the discussion of the one-loop corrections and suffice to describe the
$\omega^8$ terms found here. The enumeration of the terms with two derivatives
can be done similarly and we find that 
\begin{eqnarray}\label{leff1}
{\cal L}_{\rm eff}^{\prime} & = & \frac{1}{945}\frac{\alpha^2}{m_e^6}\left[
\left(\partial^{\alpha}\partial_{\beta}F_{\mu\nu}\right)F^{\mu\nu}
F_{\alpha\lambda}F^{\lambda\beta} +
3\left(\partial^{\alpha}F_{\mu\nu}\right)\left(\partial_{\alpha}F^{\mu\nu}
\right)F_{\lambda\rho}F^{\lambda\rho}\right. \nonumber \\
&   &\left.\hspace{54pt} + 11\left(\partial^{\alpha}F_{\mu\nu}\right)
F^{\nu\lambda}\left(\partial_{\alpha}F_{\lambda\rho}\right)F^{\rho\mu}\right]
\,.
\end{eqnarray}
gives the $\omega^6$ correction and
\begin{eqnarray}\label{leff2}
{\cal L}_{\rm eff}^{\prime\prime} & = &-\frac{1}{9450}\frac{\alpha^2}{m_e^8}
\left[33\,\left(\partial_{\alpha}F_{\mu\nu}\right)\left(\partial^
{\alpha}F^{\mu\nu}\right)\left(\partial_{\beta}F_{\lambda\rho}\right)
\left(\partial^{\beta}F^{\lambda\rho}\right)  \right. \nonumber \\
&   &\left.- 106\left(\partial_{\alpha}
F_{\mu\nu}\right)\left(\partial^{\beta}F^{\mu\nu}\right)\left(\partial^{\alpha}
F_{\lambda\rho}\right)\left(\partial_{\beta}F^{\lambda\rho}\right)
 + 262\,\left(\partial^{\alpha}\partial_{\beta}
F_{\mu\lambda}\right)F^{\lambda\nu}\left(\partial^{\mu}\partial_{\nu}
F_{\alpha\rho}\right)F^{\rho\beta}\right].
\end{eqnarray}
gives the $\omega^8$ correction.

The expansion of the photon-photon scattering amplitude can thus be obtained by
computing the matrix element of the local effective Lagrangian 
${\cal L}_{\rm eff} + {\cal L}_{\rm eff}^{\prime} + {\cal L}_{\rm
eff}^{\prime\prime}$, and it is accurate for photon energies very close to the
energy scale $m_e$ \cite{rav}.

\section{Discussion}

By using the Euler-Heisenberg Largangian, Eq.\,(\ref{E-H}), to the one-loop 
level in perturbation theory we obtained an expansion of the low energy
photon-photon scattering amplitude containing terms of order $\omega^4/m_e^4$
and $\omega^8/m_e^8$. Since there is no contribution of order $\omega^6/m_e^6$, 
it is not necessary to introduce a counterterm to ensure renormalizability 
in this order. This situation is not peculiar to the one-loop approximation.
An $\omega^6/m_e^6$ correction cannot be generated by retaining more terms in 
the Euler-Heisenberg expansion or by including perturbative corrections 
with more than one loop. While we do obtain finite one-loop terms of
order $\omega^8/m_e^8$, these are higher order in $\alpha$ 
and thus should be negligible numerically. In fact, the 
discussion of the order $\omega^8/m_e^8$ is essentially complete.
The only other potential $\omega^8$
correction to the elastic amplitude is associated with term in the
Euler-Heisenberg expansion which consists of products of eight field tensors,
but this contribution also vanishes by dimensional regulation. 

This situation should be compared with the result of expanding the exact QED
amplitude in powers of $\omega^2/m_e^2$. Here, one obtains, as expected, the
same $\omega^4/m_e^4$ term, as well as an $\omega^6/m_e^6$ term and an
$\omega^8/m_e^8$ correction which is not suppressed by additional powers of
$\alpha$. 

The existence of a leading order (in $\alpha$) $\omega^8/m_e^8$ correction in
the QED expansion can be accommodated in the effective field theory approach by
an appropriate choice of the renormalized couplings $a_i$ of 
Eq.\,(\ref{lcount}).
There is still a difference between the energy dependence derived from QED and 
that implied by  simply requiring the renormalizability of the effective 
Lagrangian approach. This fact does not seem to be in conflict with the 
spirit of effective field theory, since it is always possible to add 
interactions such as
those in Eq.\,(\ref{leff1}) to recover the $\omega^6$ dependence. One cannot,
however, restrict the number of such terms by appealing to renormalizability. 
\acknowledgements

We thank Vic Teplitz, Steven Weinberg and Scott Willenbrock for discussions. 
This research was supported in part by the
U. S. Department of Energy under Grant No. DE-FG02-95ER40896,
DE-FG013-93ER40757, in part by the National Science Foundation under Grant No.
PHY-93-07780 and in part by the University of Wisconsin Research Committee with
funds granted by the Wisconsin Alumni Research Foundation.

\begin{table}
\begin{tabular}{cccc}
      & $++++$ & $++--$ & $+-+-$ \\ \hline
\mbox{\rule[-4pt]{0pt}{18pt}}$T^s_{\lambda_1\lambda_2\lambda_3\lambda_4}$ & 
$\case{8}{5}\,(1459 + 3z^2)$  & $-1760$  &$\case{968}{5}\,(1 + z)^2$ \\
\mbox{\rule[-4pt]{0pt}{18pt}}$T^t_{\lambda_1\lambda_2\lambda_3\lambda_4}$ & 
$\case{968}{5}\,(1 - z)^2$ &$-110\,(1 - z)^4$&$\case{242}{5}\,
(1 - z^2)^2$ \\
\mbox{\rule[-4pt]{0pt}{18pt}}$T^u_{\lambda_1\lambda_2\lambda_3\lambda_4}$ & 
$\case{968}{5}\,(1 + z)^2$ &$-110\,(1 + z)^4$&$\case{1}{5}\,(1 + z)^2
(743 + 1450z + 731z^2)$ \\ [4pt]
\hline
\mbox{\rule[-4pt]{0pt}{18pt}}
$\sum_{s,t,u}\,T^{i}_{\lambda_1\lambda_2\lambda_3\lambda_4}$ &
$\case{8}{5}\,(1701 + 245z^2)$ &$-220\,
(3 + z^2)^2$ &$\case{1}{5}(1 + z)^2\,(1953 + 966z + 973z^2)$
\end{tabular}
\vspace{8pt}
\caption{The contributions to the helicity amplitudes of Eq.\,(\ref{amps}) from
the Euler-Heisenberg Lagrangian used to one-loop are given in the first three
rows. In addition, $T^s_{+--+}(z) = T^s_{+-+-}(-z)$ while $T^t_{+--+}(z) = 
T^u_{+-+-}(-z)$, $T^u_{+--+}(z) = T^t_{+-+-}(-z)$ and $T^i_{+++-}(z) = 
T^i_{++-+}(z) = 0$ for $i = s, t, u$. The last row is the sum needed in 
Eq.\,(\ref{renamp}).}
\end{table}

\begin{table}
\begin{tabular}{ccccc}
      & $++++$ & $++--$ & $+-+-$ &$+++-$ \\ \hline
\mbox{\rule[-4pt]{0pt}{18pt}}$T^4_{\lambda_1\lambda_2\lambda_3\lambda_4}$ &
$-\case{176}{45}$ &$\case{8}{15}(3 + z^2)$ &$-\case{44}{45}\,(1 + z)^2$ & 0 \\
\mbox{\rule[-4pt]{0pt}{18pt}}$T^6_{\lambda_1\lambda_2\lambda_3\lambda_4}$ &
$-\case{768}{945}$ &$\case{480}{945}\,(1 - z^2)$ &$\case{96}{945}\,(1 + z)^3$ &$
\case{48}{945}(1 - z^2)$ \\
\mbox{\rule[-4pt]{0pt}{18pt}}$T^8_{\lambda_1\lambda_2\lambda_3\lambda_4}$ &
$-\case{64}{4725}\,(157 + 25\,z^2)$ &$\case{640}{4725}\,(3 + z^2)^2$&
$-\case{8}{4725}\,(1 + z)^2(191 + 82z + 91z^2)$ & 0 
\end{tabular}
\vspace{8pt}
\caption{The angular dependence of the helicity amplitude expansion in powers of
$\alpha^2(\omega/m_e)^i,\,i = 4,6,8$\,, is shown. In all cases, $T^i_{+--+}(z) =
T^i_{+-+-}(-z)$ and $T^i_{++-+}(z) = T^i_{+++-}(z)$.}
\end{table}

\begin{table}[h]
\begin{center}
\begin{tabular}{cccc}
$\omega$&	  $\sigma$	&  $R_1$	& $R_2$  \\ \hline
\mbox{\rule[-4pt]{0pt}{18pt}}0.25 &$3.20\times10^{-35}$~&0.983~& 1.001 \\
0.35	&	$2.46\times10^{-34}$ 	&	 0.966	&	 1.010 \\
0.45	&	$1.15\times10^{-34}$		&	 0.931	&
1.020 \\
0.55	&	$4.08\times10^{-33}$		&	 0.876	&
1.036 \\
0.65	&	$1.22\times10^{-32}$	 &  0.799	&	 1.054 \\
0.75	&	$3.32\times10^{-32}$	 &	 0.692 &	 1.054  \\
0.85	&	$9.07\times10^{-32}$	 &	 0.537 &	 0.977  \\
0.95	&	$3.02\times 10^{-31}$ &  0.314	 &  0.701 \\
0.99	&	$6.74\times 10^{-31}$	 &	 0.180		&
0.438 \\
1.00 & $1.26\times 10^{-30}$	&	 0.102		&	0.255 \\
\end{tabular}
\end{center}
\caption{The first column gives the photon center-of-mass energy in units of the
electron mass.  The second column is the full one loop cross section, as 
calculated numerically, in cm$^2$.  This is the same cross section as in 
Fig. 1.  The third column is the ratio of the approximate cross section (3) 
to the full cross section in column 2.  The fourth column is the ratio of the 
improved cross section (7) to the full cross section.}
\end{table}

\begin{figure}[h]
\hspace{0.9in}
\epsfysize=2.1in \epsfbox{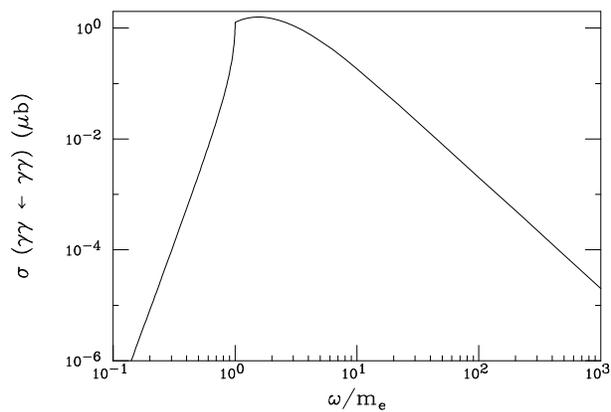}
\caption{The exact cross section for $\gamma\gamma\protect\rightarrow
\gamma\gamma$ is shown.}
\end{figure}

\begin{figure}[h]
\hspace{1.25in}
\epsfysize=2.1in \epsfbox{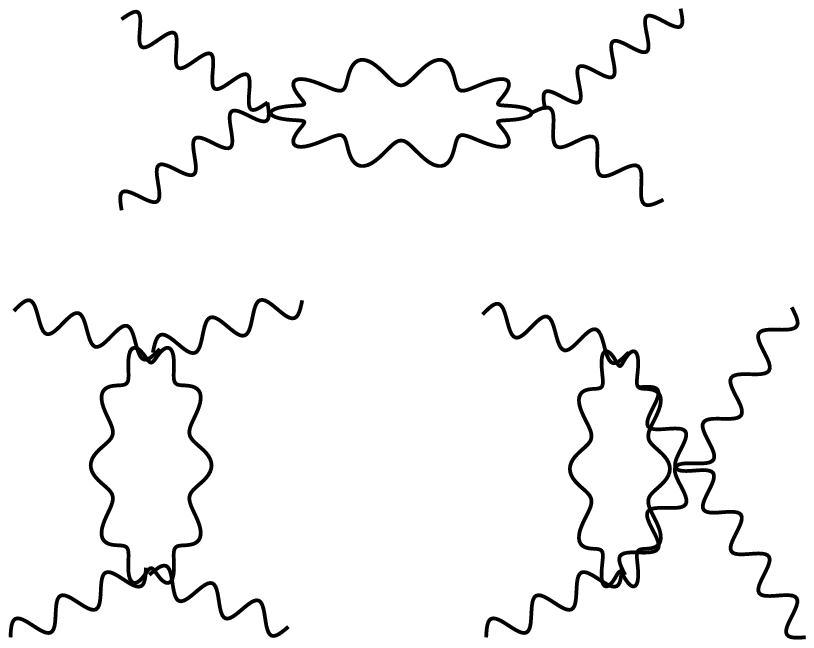}
\vspace{10pt}
\caption{One-loop diagrams for the $s$, $t$ and $u$ channel two photon 
exchanges are shown.}
\end{figure}


\begin{thebibliography}{99}

\bibitem{Eul} H. Euler, Ann. Phys. (Leipzig) {\bf 26}, 398 (1936); W. Heisenberg
and H. Euler, Zeit. Phys. {\bf 98}, 714 (1936).
\bibitem{wein} S. Weinberg, Physica {\bf 96A}, 327 (1979); Proc. of the XXVI
International Conference on High Energy Physics, edited by J. R. Sanford, AIP
Press (1993), p. 352; University of Texas Theory Group report UTTG-05-97, 
hep-th/9702027 (1997).
\bibitem{Kar} R. Karplus and M. Neuman, Phys. Rev. {\bf 80}, 380 (1950).
\bibitem{DeT} B. DeTollis, Nuovo Cim. {\bf 35}, 1182 (1965).
\bibitem{2to4} In this paper, we consider only elastic scattering,
$\gamma\gamma\rightarrow\gamma\gamma$. The second term in Eq.\,(\ref{E-H}) gives
the low energy ($\omega\,<\,m_e$) contribution to $\gamma\gamma\rightarrow
4\gamma$. The total cross section for the inelastic
process $\gamma\gamma\rightarrow 4\gamma$ is $\sigma^{(6)} =
8.0\times 10^{-40}(\omega/m_e)^{14}$\,cm$^2$. This can be compared to Eq.\,
(\ref{xsec4}), which gives $\sigma^{(4)} = 1.3\times 10^{-31}(\omega/m_e)^6$\,
cm$^2$.
\bibitem{hal} A first try along these lines was made by J. Halter, 
Phys. Lett. B {\bf 316}, 155 (1993). Note that the vertex function ${\cal
M}^{\alpha\beta\mu\nu}$ given in the Appendix contains a misprint: the first
terms in lines 7, 9, 11, 13, 15 and 17 should not be multiplied by 2.
\bibitem{velt} M. J. G. Veltman, {\it SCHOONSCHIP A Program for Symbol
Handling}, University of Michigan, report, 1984 (unpublished).
\bibitem{verm} J. A. M. Vermaseren, {\it The Symbolic manipulation program
FORM}, Report No. KEK-TH-326, 1992 (unpublished).
\bibitem{Kao} C. Kao and D.A. Dicus, LOOP, a Fortran program for evaluating loop
integrals based on the following two papers: G. t'Hooft and M. Veltman, Nuc.
Phys B {\bf 153}, 365 (1979), and G. Passarino and M. Veltman, Nuc. Phys. B {\bf
160}, 151 (1979).
\bibitem{rav} There are also contributions of 
order $\alpha^5\omega^4/m_e^4$ arising from the one-loop corrections to the box
diagram.  See: F. Ravndal, {\em Applications of Effective Lagrangians},
hep-ph/9708449; V. I. Ritus, Proc. Lebed. Phys. Inst. {\bf 168}, Nova Science
Pub., New York, N.Y. (1986); M. Reuter, M. G. Schmidt and C. Schubert,
hep-th/9610191 (1996). This effect multiplies Eqs.\,(\ref{sig0}), (\ref{xsec4}) 
and the first term of (\ref{xsec6}) by $[1 + (186685/17514)(\alpha/\pi)]$. 
The corrections we discuss exceed this correction when $\omega/m_e > 0.28$.

\end{thebibliography}
\end{document}